\documentclass[intlimits,twoside,a4paper]{article}

\usepackage{amsmath} 
\usepackage[cp1251]{inputenc}

\usepackage[eqsecnum]{cmpj3}

\issue{2024}{27}{1}{13607}
\doinumber{10.5488/CMP.27.13607}

\title[Dimerizing hard spherocylinders in porous media]%
{Dimerizing hard spherocylinders in porous media}
\author[V. I. Shmotolokha,  M. F. Holovko]{V. I. Shmotolokha\orcid{0000-0003-2729-4777}\thanks{Corresponding author: \email{shmotolokha@icmp.lviv.ua}.},
	M. F. Holovko\orcid{0000-0001-8114-5356}}
\address{
 Institute for Condensed Matter Physics of the National Academy of Sciences of Ukraine, 1 Svientsitskii St., 79011 Lviv, Ukraine
}


\date{Received November 21, 2023}

\begin{document}

\maketitle

\begin{abstract}
	This research focuses on the unique phase behavior of non-spherical patchy colloids in porous environments. Based on the theory of scaled particle (SPT), methods have been refined and applied to analyze the thermodynamic properties of non-spherical patchy particles in a disordered porous medium. Utilizing the associative theory of liquids in conjunction with SPT, we investigated the impact of associative interactions and connections between the functional nodes of particles on the formation of the nematic phase. Calculations of orientational and spatial distributions were conducted, which helped to understand the phase behavior of particles during the transition from isotropic to nematic phase under the spatial constraints imposed by the disordered matrix of the porous medium.

	\printkeywords
	%
\end{abstract}

\section{Introduction}
It is a big pleasure for us to dedicate this paper to our good friend and colleague Jaroslav Ilnytskyi for his 60th birthday celebration. His work on computer simulations of liquid crystals and polymers is very important in the physics of soft matter. Some of his studies, see, e.g., \cite{IlSokPiz,EarIlWil,IlWil,IlTrokhSch}, are methodologically related to the objectives of the present paper, specifically to computer simulations of isotropic-nematic phase transition in different molecular models, in particular in the presence of porous media or colloidal particles. M.H. is very happy that together with Jaroslav and other colleagues he was the coauthor of a common paper \cite{IlPatsHolKrMak} devoted to the development of the dissipative particle dynamics approach for the study of morphological changes in block copolymer melts due to variation of intramolecular branching.

During the last decades, the study of complex colloidal particles with chemically or physically patterned surfaces, commonly referred to as patchy colloids, has taken significant attention~\cite{BiBlLik,KalBiFerKah}. Usually, such patchy colloids are modelled as hard spheres carrying a finite small number of attractive sites arranged in precise geometries on the particle’s surface. The anisotropy and saturation of the interaction between such particles lead to a wide range of physical phenomena. The patchy colloidal system was considered to be confined in a random porous media. The influence of porous media on the phase behaviour, percolation, and dynamical properties of confined patchy colloidal fluids were considered \cite{KalyuzhnyiHolovko2014,Kor28}.

Recently, in the theory of patchy particles, along with the study of spherical colloids, the study of non-spherical patchy colloids has also become of considerable interest. The non-spherical shape of colloidal particles significantly enriches the properties of such colloidal systems. In particular, due to the non-spherical shape of colloids, there may be an orientational ordering of colloids and various associated liquid crystal phases, which can be significantly enhanced and modified by additional associative interactions due to the patchiness of colloids. Examples of such anisotropic patchy colloids are self-associated structures that occur in micellar systems \cite{1_KunWal2008}, in fibril formations \cite{2_Lee2009}, in solutions of DNA molecules~\cite{3_SauLanJan2017}, and in inorganic nanoparticles \cite{LiZhKum}. The formation of associated clusters in non-spherical patchy colloids is often considered as the major reason for much of the interesting behaviors and properties of liquid crystal systems. For example, association in liquid crystals has been considered as a possible cause of reentrant phenomena in which less ordered phases reappear with decreasing temperature \cite{Cladis}.


There are two important features inherent to non-spherical patchy colloids. The first one is connected with the rigid non-spherical core of colloidal particles, which leads to the formation of different orientational ordered liquid crystal phases, the simplest of which is the transition between non-ordered and orientationally ordered phases, known as the isotropic-nematic phase transition. The second feature is connected with the decoration of colloidal particles by attractive sites, which leads to bonding between colloidal particles. In this paper, we consider the simplest model of patchy colloids of non-spherical shape as a system of spherocylinders with one sticky point at one end of the spherocylinder that simulates the formation of chemical or hydrogen bonds between colloids. Due to this bond, the colloids can form dimers of two antiparallel spherocylinders attached end-to-end. For the first time, such a simple patchy colloidal model was introduced by Sear and Jackson \cite{5_SearJac1994}. In their consideration, they combined the Onsager theory \cite{Onsager} in order to account for the effect of an anisotropic phase on the excluded volume interactions between spherocylinders, and Wertheim's multidensity formalism \cite{CM_2015_54,CM_2015_54_a} for the treatment of the effect of dimerization of spherocylinders. If the bonding is strong enough, the hard spherocylinders will all dimerize, and the phase behavior of dimerizing hard spherocylinders will be approximately the same as for the hard spherocylinders of twice the length. As a result, it was shown \cite{5_SearJac1994} that the nematic phase is stabilized by dimerization relative to the isotropic phase. However, it should be noted that the Onsager theory is based on the low-density expansion of the free energy functional truncated at the second virial coefficient level and is exact only for a very specific model of a hard spherocylinder fluid in which the length of the spherocylinder $L_{1}\rightarrow\infty$ and the diameter $D_{1}\rightarrow 0$ in such a way that the non-dimensional density of the fluid $c_{1}=\frac{1}{4} \piup L_{1}^{2}D_{1}$ is fixed, where $\rho_{1}=N_{1}/V$, $N_{1}$ is the number of spherocylinders, and $V$ is the volume of the system \cite{VrLek}. In order to incorporate the higher-order contributions neglected in the Onsager theory, in the next study \cite{6_McSearJac1997} the authors used the Parsons--Lee (PL) approach~\cite{Parsons79,Lee87}, which is based on the mapping of the properties of the hard spherocylinder fluid to those of the hard sphere fluid. The accuracy of the developed theory was demonstrated by the comparison with original Monte-Carlo simulation data for the same model.

In the present manuscript, we study the influence of random porous media on the properties of dimerizing hard spherocylinders. Random porous media are usually considered as quenched configurations of randomly distributed obstacles \cite{MadGl}. Similar to the bulk case, the description of dimerizing hard spherocylinders in random porous media needs to combine the Onsager theory and Wertheim's multidensity approach and generalize them for the patchy colloidal system in a random porous media. However, in contrast to the bulk case, the application of the Parsons--Lee formalism for the improvement of Onsager theory in the presence of porous media seems to be quite problematic because the Parsons--Lee approach significantly uses the virial equation of state for the hard sphere fluid, the generalization of which for the presence of porous media is currently impossible in an analytical form. Instead of the Parsons--Lee approach, in this paper, we use the scaled particle theory (SPT) previously developed for a hard-sphere fluid \cite{ReiFrLeb} and during the last decade extended to generalize for the description of a hard sphere fluid in disordered porous media \cite{HolDong,ChenDongHol,HolShmotDong,PatsahanHolovko2011,HolPatDong,HolovkoPatsahan2013,ChenZhHolDong,HolPatsDong,HolKor}. The approach proposed in \cite{HolDong}, referred to as SPT1, contains a subtle inconsistency manifested when the size of matrix particles is considerably larger than the size of fluid species. This inconsistency was eliminated in a new version labeled as SPT2 \cite{PatsahanHolovko2011}. As a result, the first rather accurate analytical expressions were obtained for the chemical potential and pressure of a hard sphere fluid in a hard sphere matrix. These expressions include three parameters describing the porosity of the matrix \cite{HolPatDong,HolovkoPatsahan2013,HolPatsDong,HolKor}. The first one is related to the bare geometry of the matrix. It is the so-called geometrical porosity, $\phi_{0}$, that characterizes the free volume not occupied by the matrix particles. The second parameter, $\phi$, is defined by the chemical potential of a fluid in the limit of infinite dilution. It is the so-called probe particle porosity, which means the probability to find a fluid particle in an empty matrix. The third parameter, $\phi^{*}$, is determined by the maximum value of the fluid packing fraction of a hard sphere fluid in a porous medium. It characterizes the maximum adsorption capacity of a matrix for a given type of fluid. We note that for thermodynamic properties of a hard sphere fluid, SPT produces the same result as the Percus--Yevick theory \cite{Th1963,Wertheim}. A basic defect of such a description is known to appear at higher densities where the theory needs some improvement, such as a semi-empirical Carnahan--Starling (CS) correction \cite{CarStar,YuHol}.

In the present manuscript, we study the influence of random porous media on the properties of dimerizing hard spherocylinders. The scaled particle theory was generalized for the hard spherocylinder fluid \cite{Cotter1974,Cotter1978} and for the mixture of hard spheres and hard spherocylinders \cite{HolHv,LagoCueMarRull} for the bulk case, in order to incorporate the higher-order contributions neglected in the Onsager theory. As a result, it became possible to generalize the Onsager theory for the description of a more realistic model of the hard spherocylinder fluid with a finite value of the spherocylinder length and a non-zero value of the diameter. However, by comparison with corresponding computer simulation data \cite{HolHv,LagoCueMarRull}, it was shown that the accuracy of the SPT description reduces with the decreasing length $L_{1}$ of spherocylinders. This reduced accuracy of the SPT description was improved by the CS correction within the framework of the Parsons--Lee approach, which leads to a correct description of the isotropic-nematic transition in the hard spherocylinder fluid, even at small lengths $L_{1}$ of spherocylinders \cite{Lee87}. However, as we have already noted, the generalization of the PL approach for the presence of porous media is problematic. For the presence of disordered porous media, the SPT2 approach was generalized for the fluid of hard convex body particles in disordered porous media \cite{HolShmotPats2014} and was used for the study of the influence of porous media on the isotropic-nematic transition in a hard spherocylinder fluid in disordered porous media~\cite{HolShmot2018,book_11} and in a hard spherocylinder-hard sphere mixture \cite{HvPatHol}. It was shown that a porous medium shifts the isotropic-nematic phase transition to smaller fluid densities. However, similar to the bulk case, the accuracy of the developed SPT description reduces with a decreasing spherocylinder length. To address this, in \cite{HolShmot2018,book_11} two types of corrections were introduced to improve the SPT description of a hard spherocylinder fluid in disordered porous media. The first one is the CS correction, which improves the description at higher fluid densities. The second one corrects the description of the orientational ordering in a hard spherocylinder fluid at higher densities. This correction was formulated by comparing the constants in the integral equation for the singlet distribution function of hard spherocylinders in the SPT approach and in the PL theory in the bulk case. The CS and PL corrections constitute the improvement of the SPT description of a hard spherocylinder fluid in disordered porous media. It was shown that both corrections provide a correct description of the isotropic-nematic phase transition in a hard spherocylinder fluid in disordered porous media, including the hard spherocylinder fluids with small lengths of spherocylinders.
The obtained results for hard spherocylinders in random porous media were used as the reference system for the generalization of the Van der Waals equation for anisotropic fluids in random porous media \cite{book_11,HolShmot2015,HolShmot}, which was used for the description of the influence of the porous media on the phase behavior of polypeptide solutions \cite{ShmotHol}. A model of hard spherocylinders-charged hard spheres mixture in random porous media was used for the description of the phase behavior of electrolyte solutions in anisotropic solvents \cite{HvPatPatHol,HvPatPatHol2022}.

In the present paper, we use the obtained results for the hard spherocylinder fluid in random porous media as the reference system for the study of the influence of porous media on the phase behavior of dimerized hard spherocylinders. We show that due to association, the phase behavior of dimerizing hard spherocylinders is strongly temperature-dependent, and a porous medium shifts the isotropic-nematic transition to smaller fluid densities and lower temperatures.
The paper is arranged as follows: In section~2, we formulate the model of dimerizing hard spherocylinders in random porous media and present the theory which combines the SPT approach for a hard spherocylinder fluid in disordered porous media with the CS and PL corrections and the associative multidensity formalism for the description of the dimerizing effects. The results and discussion are presented in section 3. We conclude in section 4.

\section{Model and theory}
\begin{figure}[h]
	\begin{center}
		\includegraphics[clip,width=0.6\textwidth,angle=0]{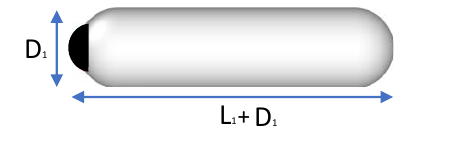}
		\caption{(Colour online) Schematic representation of the model dimerizing hard spherocylinder.}
		\label{Fig10}
	\end{center}
\end{figure}

In the present study, we examine the qualitative features of the phase behavior of dimerized hard spherocylinder fluid in disordered porous media. To enable a comparison of our theoretical results with the computer simulation data obtained in \cite{6_McSearJac1997} for the bulk case, we consider here the dimerizing hard spherocylinder model with the same parameters as in \cite{6_McSearJac1997}. The schematic representation of the model dimerizing hard spherocylinder is presented in figure~\ref{Fig10}. The model is a hard spherocylinder with a cylindrical core of length $L_{1}$, capped at each end by hemispheres with a diameter $D_{1}$, and with a square-well bonding site embedded in one of the hemispherical end caps at a distance $r_\text{d} =0.25D_{1}$ from the center of the hemisphere to the surface along the molecular axis, as shown in figure~\ref{Fig10}. The bonding sites mediate the association of the dimerizing spherocylinders, described by the site-site square-well potential.

\begin{eqnarray}
	\phi_{\rm attr}\left(r\right)=\left\{\begin{array}{ll}
		-\epsilon_{1}, & r \leqslant \delta_{1}, \\
		0, & r \geqslant  \delta_{1},
	\end{array}\right.
	\label{hol2.38}
\end{eqnarray}
where $r$ is the site-site distance between the bonding sites of two spherocylinders. The parameter $\epsilon_{1}$ represents the depth of the square-well and is used to define the dimensionless temperature $T^{*} = kT/\epsilon_{1}$, where $k$ is the Boltzmann constant and $T$ is the temperature on the absolute Kelvin scale. The parameter $\delta_{1}$ denotes the width of the square-well and satisfies the following inequality \cite{KalyuzhnyiHolovko2014}: $\delta_{1} \leqslant (1-\sqrt{3}/2)D_{1} = 0.134D_{1}$, which ensures that only one attractive bond per site between spherocylinders can be formed. In the  study \cite{6_McSearJac1997}, the authors chose $\delta_{1} = 0.5742D_{1}$. In the present paper we use a similar value of the potential parameter $\delta_{1}$. That is why the bonding in the considered case was not saturable.
 However, it was shown by computer simulations that in the considered case, for densities close to the isotropic-nematic phase transition, the formation of trimers and larger aggregates is extremely rare. Therefore, in the present theoretical study, we take into account only dimerizing clusters.

We also consider that the model at hand is confined in a porous medium represented by a matrix of frozen hard spheres (HS), characterized by hard-sphere obstacles of diameter $D_{0}$, and two types of porosity, \cite{PatsahanHolovko2011, HolovkoPatsahan2013} i.e., geometrical porosity $\phi_{0}$ and probe particle porosity $\phi$. For the HS matrix, $\phi_{0} = 1-\eta_{0}$, where $\eta_{0} = \piup \rho_{0} D_{0}^3/6$ and $\phi$ is defined by the excess value of the chemical potential $\mu_{1}^{\rm ex}$ of the fluid particles in the limit of infinite dilution, i.e., $\phi = \exp(-\beta\mu_{1}^{\rm ex})$, where $\beta = 1/kT$.

As we already noted in the introductory part of the present manuscript, we combine the scaled particle theory, generalized in \cite{HolShmot2018,book_11} for the case of hard spherocylinders in porous media, with the multidensity formalism from the theory of associative fluids \cite{5_SearJac1994,CM_2015_54,CM_2015_54_a,CM_2015_55,CM_2015_55_a,13_Hol_2002}, in order to describe the associative interactions. According to the multidensity formalism, the singlet density distribution function $\rho_{1}(1)$ can be presented as the sum:
\begin{equation}
	\rho_{1}\left(1\right) = \rho_{10}\left(1\right) + \rho_{11}\left(1\right),
	\label{hol2.40}
\end{equation}
where $\rho_{10}(1)$ and $\rho_{11}(1)$ represent the singlet density distribution functions of monomers (nonbonding part) and dimers (bonding part), respectively. We note that we use the conventional notations in the theory of fluids in porous media \cite{PatsahanHolovko2011,HolovkoPatsahan2013,book_11}: the index ``1'' is used to denote the fluid component and the index ``0'' to denote the matrix particle. We also use the index ``s'' for the scaled particle and follow the conventional notations in the theory of associative fluids \cite{CM_2015_54,CM_2015_54_a,CM_2015_55,CM_2015_55_a,13_Hol_2002}. In the case of two indexes, the second index ``0'' corresponds to the nonbonded part and the second index ``1'' to the bonded part. In~(\ref{hol2.40}), we also use conventional notations in fluid theory \cite{YuHol}. The notation $(1)=(r_{1},\Omega_{1})$ denotes the coordinates (the positions $r_{1}$ and orientations $\Omega_{1}$) of a fixed spherocylinder. Here, $\Omega=(\theta,\varphi)$ represents the orientation of a particle determined by the angles $\theta$ and $\varphi$. For a homogeneous nematic phase, the singlet distribution functions $\rho_{1}(1)$ and $\rho_{10}(1)$ are functions only of the angle between the molecular axis and the preferred direction, the so-called nematic director. Thus,
\begin{align}
	\rho_{1}(1) &= \rho_{1} f_{1}(\Omega_{1}) = \rho_{1} f_{1}(\theta_{1}), \nonumber \\
	\rho_{10}(1) &= \rho_{10} f_{10}(\Omega_{1}) = \rho_{10} f_{10}(\theta_{1}),
	\label{hol2.39}
\end{align}
where $\rho_{1}$ is the total number density of spherocylinders, $\rho_{10}$ is the number density of monomers, and $\theta$ is the angle between the molecular axis and the nematic director. The functions $f_{1}(\Omega)$ and $f_{10}(\Omega)$ are normalized to unity:
\begin{equation}
\int f_{1}(\Omega)\,\rd\Omega=\int f_{10}(\Omega)\,\rd\Omega =1.
\label{hol2.39a}
\end{equation}
Here, $\rd\Omega=\frac{1}{4\piup} \sin\theta \,\rd\theta \,\rd\varphi$ is the normalized element of body angle.

For the isotropic phase, $f_{1}(\Omega)=f_{10}(\Omega)=1$.
The free energy of the system within the approximation obtained in the framework of the first-order thermodynamic perturbation theory \cite{5_SearJac1994,6_McSearJac1997,CM_2015_55,CM_2015_55_a,13_Hol_2002,Holovko1999} can be written as the sum of three contributions:
%
\begin{eqnarray}
	\beta A &=& \int \left\{\rho_{1}\left(1\right)\ln\big[\rho_{10}\left(1\right)\Lambda_{1}^{3}\big] - \rho_{10}\left(1\right)\right\}\rd 1 + \beta A_{\text{HSC}} \nonumber \\
	&-& \frac{1}{2}\int\rho_{10}\left(1\right)\rho_{10}\left(2\right)g_{\text{HSC}}\left(12\right)f_{\text{as}}\left(12\right)\rd 1 \,\rd 2,
	\label{eq:15}
\end{eqnarray}
where $\Lambda_{1}$ is the de Broglie wavelength, $g_{\text{HSC}}(12)$ is the binary distribution function of the hard spherocylinders.
The first term represents the ideal contribution to the free energy of the system of dimerizing spherocylinders. It includes the contribution from orientational entropy
\begin{equation}
	\sigma(f_{1},f_{10})=\int f_{1}(\Omega) \ln f_{10}(\Omega)\rd\Omega,
	\label{hol2.41}
\end{equation}
which appears from the first term in \eqref{eq:15} after substitution of \eqref{hol2.39}.

The second term in (\ref{eq:15}) $\beta A_{\text{HSC}}$ accounts for the hard-particle interaction of the spherocylinders, which in the Onsager limit reduces to the excluded volume integral in the form of the second virial coefficient \cite{Onsager,VrLek}
$$B_{2}=\frac{1}{2} \int f_{1}(\Omega_{1}) f_{1}(\Omega_{2}) V_{\rm exc}(\Omega_{1},\Omega_{2})\, \rd\Omega_{1}\, \rd\Omega_{2},$$
where
 $$V_{\rm exc}(\Omega_{1},\Omega_{2})=2L_{1}^{2}D_{1}\sin(\theta_{12})+2\piup L_{1}D_{1}^{2}+\frac{4}{3} \piup D_{1}^{3},$$  
$\theta_{12}$ is the  angle between the principal  axis of two  spherocylinders. 
The third term in (\ref{eq:15}) reflects the associative dimerization contribution.

\subsection{Hard spherocylinder contribution}
To describe the second term --- the hard spherocylinder contribution --- the scaled particle method is employed. This method involves introducing an additional spherocylindrical particle into the system of hard spherocylinders in porous media, characterized by two scale parameters: diameter $D_{s}$ and length of the cylinder $L_{s}$:
\begin{equation}
	D_{s} = \lambda_{s}D_{1}, \quad L_{s} = \alpha_{s}L_{1}.
	\label{eq:Hol4}
\end{equation}
Between different approximations considered in the framework of SPT2 approach we restrict here to the SPT2b1 approximation which is quite accurate at small, intermediate and higher fluid densities.
The expressions obtained using the scaled particle method for the pressure and chemical potential within the SPT2b1 approximation can be represented as follows \cite{HolShmot2018,book_11}:
\begin{eqnarray}
	\left[\beta\big(\mu_{1}^{\text{ex}}-\mu_{1}^{0}\big)\right]^{\text{SPT2b1}} &=& \sigma(f_{1},f_{10}) - \ln(1-\eta_{1}/\phi_{0}) + \big[1+A\big(\tau(f_{1})\big)\big]\frac{\eta_{1}/\phi_{0}}{1-\eta_{1}/\phi_{0}}
	\nonumber \\ &+& \frac{\eta_{1}(\phi_{0}-\phi)}{\phi_{0}\phi(1-\eta_{1}/\phi_{0})}
	+ \frac{1}{2}\big[A\big(\tau(f_{1})\big)+2B\big(\tau(f_{1})\big)\big]\frac{(\eta_{1}/\phi_{0})^{2}}{(1-\eta_{1}/\phi_{0})^{2}} \nonumber \\ &+& \frac{2}{3}B\big(\tau(f_{1})\big)\frac{(\eta_{1}/\phi_{0})^{3}}
	{(1-\eta_{1}/\phi_{0})^{3}}\,,
	\label{hol2.19}
\end{eqnarray}
\begin{eqnarray}
	\left(\frac{\beta P}{\rho_{1}}\right)^{\text{SPT2b1}} &=& \frac{1}{1-\eta_{1}/\phi_{0}}\frac{\phi_{0}}{\phi} + \left(\frac{\phi_{0}}{\phi}-1\right)
	\frac{\phi_{0}}{\eta_{1}}\ln(1-\eta_{1}/\phi_{0})\nonumber \\
	&+& \frac{A\big(\tau(f_{1})\big)}{2}\frac{\eta_{1}/\phi_{0}}{(1-\eta_{1}/\phi_{0})^{2}} + \frac{2B\big(\tau(f_{1})\big)}{3}\frac{(\eta_{1}/\phi_{0})^{2}}{(1-\eta_{1}/\phi_{0})^{3}}\,,
	\label{hol2.20}
\end{eqnarray}
where $\eta_{1} = \rho_{1} V_{1}$ is the packing fraction parameter of the system of spherocylindrical particles, $\rho_{1}$ is the density
of spherocylindrical particles, and $V_{1}=({\piup D_{1}^3}/{6}) + ({\piup L_{1} D_{1}^2}/{4})$ is the volume of the hard spherocylindrical particles.
We note that in (\ref{hol2.19}) we introduced the orientational entropy term $\sigma(f_{1},f_{10})$, defined by (\ref{hol2.41}).
The coefficients $A(\tau(f_{1}))$ and $B(\tau(f_{1}))$ are defined by the expressions.
%
%
\begin{eqnarray}
	A(\tau(f_{1})) &=& 6 + \frac{6(\gamma_{1}-1)^2\tau(f_{1})}{3\gamma_{1}-1} \nonumber \\
	&-& \frac{p'_{0\lambda}}{\phi_0}\left[4 + \frac{3(\gamma_{1}-1)^2\tau(f_{1})}{3\gamma_{1}-1}\right] \nonumber \\
	&-& \frac{p'_{0\alpha}}{\phi_0}\left(1 + \frac{6\gamma_{1}}{3\gamma_{1}-1}\right) \nonumber \\
	&-& \frac{p''_{0\alpha\lambda}}{\phi_0} - \frac{1}{2} \frac{p''_{0\lambda\lambda}}{\phi_0} + 2\frac{p'_{0\alpha}p'_{0\lambda}}{\phi_0^{2}} + \left(\frac{p'_{0\lambda}}{\phi_0}\right)^2,\\
	\label{eq:Hol15}
%
	B(\tau(f_{1})) &=& \left(\frac{6\gamma_{1}}{3\gamma_{1}-1} - \frac{p'_{0\lambda}}{\phi_0}\right) \nonumber \\
	&\times& \left[\frac{3(2\gamma_{1}-1)}{3\gamma_{1}-1} + \frac{3(\gamma_{1}-1)^2\tau(f_{1})}{3\gamma_{1}-1} - \frac{p'_{0\alpha}}{\phi_0} - \frac{1}{2}\frac{p'_{0\lambda}}{\phi_0}\right],
	\label{eq:Hol16}
\end{eqnarray}
where
\begin{equation}
	\tau(f_{1}) = \frac{4}{\piup}\int f_{1}(\Omega_{1}) f_{1}(\Omega_{2}) \sin \vartheta_{12}\,\rd\Omega_{1} \,\rd\Omega_{2}.
	\label{eq:Hol17}
\end{equation}
$p'_{0\lambda} = \frac{\partial p_{0}(\alpha_s,\lambda_s)}{\partial \lambda_s}$,
$p'_{0\alpha} = \frac{\partial p_{0}(\alpha_s, \lambda_s)}{\partial \alpha_s}$,
$p''_{0\alpha\lambda} = \frac{\partial^{2} p_{0}(\alpha_s, \lambda_s)}{\partial \alpha_s \partial \lambda_s}$,
$p''_{0\lambda\lambda} = \frac{\partial^{2} p_{0}(\alpha_s, \lambda_s)}{\partial \lambda_{s}^{2}}$ are suitable derivatives at $\alpha_{s} = \lambda_{s} = 0$.
\begin{equation}
	p_{0}(\alpha_{s}, \lambda_{s}) = \exp \big[-\beta \mu_{s}^{0}(\alpha_{s}, \lambda_{s})\big]
	\label{eq:Hol6}
\end{equation}
is the probability of finding a cavity in an empty matrix which is determined by the excess of the chemical potential $\mu_{s}^{0}(\alpha_{s}, \lambda_{s})$ of the scale particle at infinite dilution.

The probability $p_{0}(\alpha_{s}, \lambda_{s})$ is associated with two types of porosity \cite{PatsahanHolovko2011, HolovkoPatsahan2013, book_11, HolShmotPats2014}. The first is geometric porosity, defined as:
\begin{eqnarray}
	\phi_{0} = p_{0}(\alpha_{s} = \lambda_{s} = 0),
	\label{eq:Hol9}
\end{eqnarray}
which describes the free volume available for fluid molecules. In the case of a hard-spherical matrix, it is given by:
\begin{equation}
	\phi_{0} = 1 - \eta_{0},
	\label{eq:Hol10}
\end{equation}
where $\eta_{0} = \frac{1}{6} \piup D_{0}^{3} \rho_{0}$ represents the packing fraction of matrix particles, $\rho_{0}$ is the density of matrix particles, and $D_{0}$ is the diameter of matrix particles.

The second type of porosity is defined at $\lambda_{s} = \alpha_{s} = 1$, which corresponds to thermodynamic porosity.
\begin{eqnarray}
	\phi = p_{0}(\alpha_{s} = \lambda_{s} = 1) = \exp(-\beta \mu_{1}^{0}),
	\label{eq:Hol11}
\end{eqnarray}
which is described by the excess of the chemical potential of the molecules of fluid $\mu_{1}^{0}$ at infinite dilution. In this case we are considering \cite{book_11}
\begin{eqnarray}
	\phi &=& (1-\eta_{0})\exp\left\{-\frac{\eta_{0}}{1-\eta_{0}}\tau\left[\frac{3}{2}(\gamma_{1}+1)+3\gamma_{1}\tau\right]\right.\nonumber\\
	&-&\left.\frac{\eta_{0}^{2}}{(1-\eta_{0})^{2}}\frac{9}{2}\gamma_{1}\tau^{2}\right.\nonumber\\
	&-&\left.\frac{1}{2}\frac{\eta_{0}}{(1-\eta_{0})^{3}}(3\gamma_{1}-1)\tau^{3}(1+\eta_{0}+\eta_{0}^{2})\right\},
	\label{eq:Hol12}
\end{eqnarray}
where $\tau = \frac{D_{1}}{D_{0}}$, $\gamma_{1} = 1 + \frac{L_{1}}{D_{1}}$.

For the free energy, we can get an expression from the thermodynamic relationship
\begin{eqnarray}
	\frac{\beta A}{V} = \beta \mu_{1}\rho_{1} - \beta P.
	\label{hol2.22}
\end{eqnarray}
The free energy, limited by fluid in the SPT2b1 approximation, is presented as follows:
\begin{eqnarray}
	 \left(\frac{\beta A}{N}\right)^{\text{SPT2b1}} &=& \sigma(f_{1},f_{10}) + \ln\frac{\eta_{1}}{\phi} - 1 - \ln(1 - \eta_{1}/\phi_{0})\nonumber\\
	&+& \left(1 - \frac{\phi_{0}}{\phi} \right)\bigg[1 + \frac{\phi_{0}}{\eta_{1}}\ln(1 - \eta_{1}/\phi_{0})\bigg] \nonumber\\
	&+& \frac{A\big(\tau(f_{1})\big)}{2} \frac{\eta_{1}/\phi_{0}}{1 - \eta_{1}/\phi_{0}} + \frac{B\big(\tau(f_{1})\big)}{3}
	\left(\frac{\eta_{1}/\phi_{0}}{1 - \eta_{1}/\phi_{0}}\right)^2.
	\label{hol2.23}
\end{eqnarray}
According to the \cite{HolShmot2018},  we should add the CS terms to expressions \eqref{hol2.19}, \eqref{hol2.20}, and \eqref{hol2.22}.
\begin{eqnarray}
\left( \frac{\beta P}{\rho_{1}} \right)^{\text{SPT2b1-CS}} = \left( \frac{\beta P}{\rho_{1}} \right)^{\text{SPT2b1}} + \left( \frac{\beta P}{\rho_{1}} \right)^{\text{CS}},
\end{eqnarray}

\begin{eqnarray}
\left(\beta \mu_{1}\right)^{\text{SPT2b1-CS}} = \left(\beta \mu_{1}\right)^{\text{SPT2b1}} + \left(\beta \Delta \mu_{1}\right)^{\text{CS}},
\end{eqnarray}
\begin{eqnarray}
\left( \frac{\beta A}{N_{1}} \right)^{\text{SPT2b1-CS}} = \left( \frac{\beta A}{N_{1}} \right)^{\text{SPT2b1}} + \left( \frac{\beta A}{N_{1}} \right)^{\text{CS}},
\end{eqnarray}
where
\begin{eqnarray}
\left( \frac{\beta \Delta P}{\rho_{1}} \right)^{\text{CS}} = -\frac{(\eta_{1} / \phi_{0})^{3}}{(1 - \eta_{1} / \phi_{0})^{3}},
\end{eqnarray}

\begin{eqnarray}
	\left(\beta \Delta \mu_{1}\right)^{\text{CS}} &=& \ln\left(1 - \frac{\eta_{1}}{\phi_{0}}\right) + \frac{\eta_{1}/\phi_{0}}{1 - \eta_{1}/\phi_{0}} - \frac{1}{2}  \left(\frac{\eta_{1}/\phi_{0}}{1 - \eta_{1}/\phi_{0}}\right)^{2} \nonumber \\
	&& - \left(\frac{\eta_{1}/\phi_{0}}{1 - \eta_{1}/\phi_{0}}\right)^{3},
\end{eqnarray}
\begin{equation}
	\left( \frac{\beta \Delta A}{N_{1}} \right)^{\text{CS}} = \ln\left(1 - \frac{\eta_{1}}{\phi_{0}}\right) + \frac{\eta_{1}/\phi_{0}}{1 - \eta_{1}/\phi_{0}} - \frac{1}{2}  \left(\frac{\eta_{1}/\phi_{0}}{1 - \eta_{1}/\phi_{0}}\right)^{2}.
\end{equation}
\newpage

\subsection{Associative contribution and the integral equations for the singlet distribution functions}

The free energy of the considered system is given by the expression (\ref{eq:15}), in which the first two terms can be presented in the form (\ref{hol2.23}). This expression includes contributions from hard spherocylinders and from associative interactions. To complete this, we should determine the expressions for the singlet distribution functions $\rho_{1}(1) = \rho_{1} f_{1}(\Omega_{1})$ and $\rho_{10}(1) = \rho_{10} f_{10}(\Omega_{1})$. Both functions can be found from the minimization of the free energy with respect to variations in these distributions. In particular, from variation in $\rho_{10}(1)$, we have the generalization of a very known relation between $\rho_{1}(1)$ and $\rho_{10}(1)$ in the theory of associative fluids for the anisotropic associative fluids in porous media, which plays the role of the mass action law (MAL) in the theory of associative fluids:
\begin{equation}
	\rho_{1}(1) = \rho_{10}(1) + \rho_{10}(1) \int \rho_{10}(2) g_{\text{HCS}}(12) f_{\text{as}}(12) \, \rd 2.
	\label{eq:2.19}
\end{equation}
In this relation, it is convenient, according to (\ref{hol2.39}), to move from $\rho_{1}(1)$ and $\rho_{10}(1)$ to functions $f_{1}(1)$ and $f_{10}(1)$.
Since in the isotropic phase the functions $f_{1}(1) = f_{10}(1) = 1$ and in the nematic phase all spherocylinders are nearly parallel, in (\ref{eq:2.19}) we can use the approximation $f_{10}(2) = f_{10}(1)$, for the first time introduced by Sear and Jackson \cite{5_SearJac1994}. Finally, due to the delta-like character of the associative interaction, the relation (\ref{eq:2.19}) can be written in the form:
\begin{equation}
	f_{1}(\Omega_{1}) = f_{10}(\Omega_{1}) X + X^{2} f_{10}^{2}(\Omega_{1}) \rho_{1} K F g_{\text{HCS}}^{\text{cont}},
	\label{eq:2.20}
\end{equation}
where $X = {\rho_{10}}/{\rho_{1}}$ is the fraction of monomers, $F = \exp(\beta \epsilon_{1}) - 1$, and $K$ is the geometric multiplier, determined by the volume of overlap of the two interactive bonding sites \cite{Chapman1988}. The contact value of the binary function of hard spherocylinders $g_{\text{HSC}}^{\text{cont}}$ will be approximated by the corresponding contact value of hard spheres, which, with the help of previous works \cite{KalyuzhnyiHolovko2014,Kor28}, can be presented in the following form:
\begin{equation}
	 g_{\text{HCS}}^{\text{cont}} = g^{\text{cont}} = \frac{1}{\phi_{0} - \eta_{1}} + \frac{3}{2}\frac{\delta\eta_{1} + \tau\eta_{0}}{(\phi_{0} - \eta_{1})^{2}} + \frac{1}{2}\frac{(\delta\eta_{1} + \tau\eta_{0})^{2}}{(\phi_{0} - \eta_{1})^{3}},
	\label{eq:2.21}
\end{equation}
where $\delta = {2\gamma_{1}}/({3\gamma_{1} - 1}), \tau = {\sigma_{1}}/{\sigma_{0}}$.

After integration of equation~(\ref{eq:2.20}) over the angles, we obtain the following equation for $X$:
\begin{equation}
	1 = X + X^{2}  \rho_{1} K F \bar{f_{10}^{2}} g^{\text{cont}},
	\label{eq:2.22}
\end{equation}
where
\begin{equation}
	\bar{f_{10}^{2}} = \int \rd\Omega f_{10}^{2}(\Omega).
	\label{eq:2.23}
\end{equation}
The second relation between $f_{1}(1)$ and $f_{10}(1)$ can be found from the minimization of the free energy with respect to $\rho_{1}(1)$:
\begin{equation}
	\ln f_{10}(\Omega_{1}) + 1 + C\frac{8}{\pi} \int f_{1}(\Omega_{2}) \sin \vartheta_{12} \, \rd\Omega_{2}=0,
	\label{eq:2.24}
\end{equation}
where
\begin{eqnarray}
	C&=& \frac{\eta_{1}/\phi_{0}}{1-\eta_{1}/\phi_{0}}\left[\frac{3(\gamma_{1}-1)^{2}}{3\gamma_{1}-1}\left(1-\frac{p'_{0\lambda}}{2\phi_{0}}\right)\right.\nonumber\\
	&+& \left.\frac{\eta_{1}/\phi_{0}}{1-\eta_{1}/\phi_{0}}\nu\frac{(\gamma_{1}-1)^{2}}{3\gamma_{1}-1}\left(\frac{6\gamma_{1}}{3\gamma_{1}-1}
	-\frac{p'_{0\lambda}}{\phi_{0}}\right)\right].
	\label{eq:2.25}
\end{eqnarray}
$\nu = \frac{3}{8}$ is the Parsons--Lee correction introduced by us in \cite{HolShmot2018}. As a result, we found that the functions $f_{1}(\Omega)$ and $f_{10}(\Omega)$ are defined by the systems of equations~(\ref{eq:2.20}) and~(\ref{eq:2.24}). Substituting from expression~(\ref{eq:2.20}) the equation for the unary function $f_{1}(\Omega_{1})$ into expression (\ref{eq:2.24}), we obtain the following equation for the function $f_{10}(\Omega_{1})$:
\begin{eqnarray}
	\ln f_{10}(\Omega_{1}) &=& -1 - C X \frac{8}{\piup} \int f_{10}(\Omega_{2})\sin\vartheta_{12}\,\rd\Omega_{2} \nonumber \\ 
	&-& \frac{8}{\piup} C X^{2}\rho_1 K F  g^{\text{cont}}\int f_{10}^{2}(\Omega_{2})\sin\vartheta_{12}\,\rd\Omega_{2}.
	\label{eq:2.26}
\end{eqnarray}
After the normalization, equation (\ref{eq:2.26}) can be represented in the form:
\begin{align}
&f_{10}\left(\Omega_{1}\right) \nonumber \\ 
&= \frac{\exp\left[-\frac{8}{\piup}CX \int f_{10}\left(\Omega_{2}\right)\sin\gamma_{12}\rd\Omega_{2}\right] \exp\left[-\frac{8}{\piup} C\rho_1 K g^{\rm cont}F X^{2}\int f_{10}^{2}\left(\Omega_{2}\right)\sin\gamma_{12}\rd\Omega_{2}\right]} {\frac{8}{\piup}\int\exp\left[-\frac{8}{\piup} CX\int f_{10}\left(\Omega_{2}\right)\sin\gamma_{23}\rd\Omega_{2}\right] \exp\left[-\frac{8}{\piup} C\rho_1 K g^{\rm cont}F X^{2}\int f_{10}^{2}\left(\Omega_{2}\right)\sin\gamma_{23}\rd\Omega_{2}\right]\rd\Omega_{3}}.
	\label{eq:2.27}
\end{align}
This equation is solved numerically by generalizing the iterative procedure proposed in \cite{HerBerWin84} for a more simple case when the associative interactions are absent. Now, the expressions for free energy, pressure, and chemical potential can be written as the sum of two terms:
\begin{equation}
	\frac{\beta A}{N_{1}} = \frac{\beta A_{\text{ref}}}{N_{1}} + \frac{\beta A_{\text{as}}}{N_{1}},
	\label{eq:2.28}
\end{equation}
\begin{equation}
	\frac{\beta P}{\rho_{1}} = \frac{\beta P^{\text{ref}}}{\rho_{1}} + \frac{\beta P^{\text{as}}}{\rho_{1}},
	\label{eq:2.29}
\end{equation}
\begin{equation}
	\beta\mu_{1} = \beta\mu_{1}^{\text{ref}} + \beta\mu_{1}^{\text{as}},
	\label{eq:2.30}
\end{equation}
where the first term has a form similar to that for hard spherocylinders in porous media, with a modification in the entropy term $\sigma(f)$ to $\sigma(f_{1}, f_{10})$, as defined by (\ref{hol2.41}). In accordance with thermodynamic perturbation theory, the associative terms can be written as follows:

\begin{equation}
	\frac{\beta A_{\text{as}}}{N_{1}} = \left(\ln X - \frac{X}{2} + \frac{1}{2}\right),
	\label{eq:31}
\end{equation}
\begin{equation}
	\frac{\beta P^{\text{as}}}{\rho_{1}} = -\frac{1}{2}\left(1 - X\right)\left(1 + \rho_{1}\frac{\partial\ln g^{\text{cont}}}{\partial\rho_{1}}\right),
	\label{eq:32}
\end{equation}
\begin{equation}
	\beta\mu_{1}^{\text{as}} = \left[\ln X - \frac{1}{2}\left(1 - X\right)\rho_{1}\frac{\partial\ln g^{\text{cont}}}{\partial\rho_{1}}\right].
	\label{eq:33}
\end{equation}

\section{Results and discussions}

\begin{figure}[h]
	\begin{center}
		\includegraphics[clip,width=0.6\textwidth,angle=0]{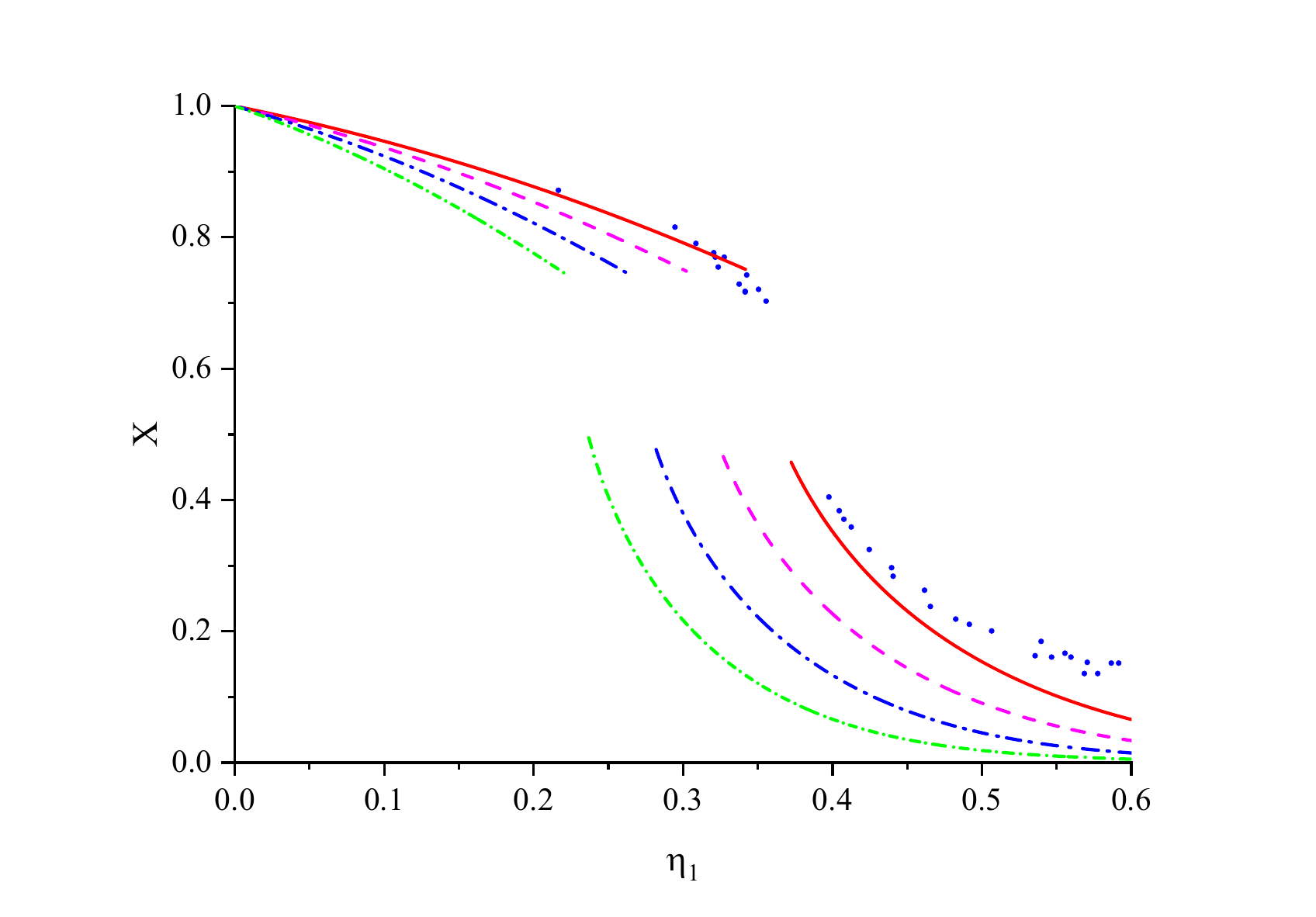}
		\caption{(Colour online) The fraction of monomers $X$ as a function of $\eta_{1}$ for the dimerization of hard spherocylinders with an aspect ratio $L_{1}/D_{1} = 5$. Data points, representing results from simulations~\cite{6_McSearJac1997}, are plotted alongside the curves derived from theoretical models. These curves correspond to the isotropic and nematic phases. The colors of the curves indicate various packing fractions of the matrix, represented by $\eta_{0} = 0$ (red color), $0.1$ (magenta color), $0.2$ (blue color), and $0.3$ (green color). }
		\label{Fig1}
	\end{center}
\end{figure}

In this section, we  illustrate the theory presented in the previous section for the dimerizing hard spherocylinder fluid in a hard sphere matrix. During this study, we aim to elucidate the influence of associative interactions and porous media on the isotropic-nematic phase transition in the system of dimerizing hard spherocylinders. It is noted that, according to the results of computer simulation \cite{BolFren1997}, in a system of hard spherocylinders, only an isotropic phase is observed at $L_{1}/D_{1} < 3.7$, and a transition from isotropic to nematic phase is observed at $L_{1}/D_{1} > 3.7$. In order to check the accuracy of our theoretical predictions, we also compare the obtained theoretical results with the data from computer modelling~\cite{6_McSearJac1997} for the dimerization model of hard spherocylinders in the bulk case. For this purpose, we consider spherocylinders with an aspect ratio of $L_{1}/D_{1} = 5$. We  also select interaction parameters between patches, as in the case of computer simulation  \cite{6_McSearJac1997}. In theory, the specific geometry of the bonding site is not provided, whereas in simulation studies, this area is strictly defined. The bonding volume of the overlap of two sites $K= 5.17 \times 10^{-4} D_{1}^{3}$, and the temperature $T^{*} = kT/\epsilon_{1} = 0.1429$. This temperature corresponds to the depth of the square-well $\epsilon_{1} = 7kT$, indicating that the formed bonds will be relatively strong. Figure~\ref{Fig1}  shows the dependence of the fraction of monomers $X$ on the packing fraction of hard spherocylinders $\eta_{1}$ at $L_{1}/D_{1} = 5$ with one sticky point site. A matrix is modelled by a frozen configuration of hard spheres randomly located in space with the packing fraction $\eta_{0} = 0.1, 0.2, 0.3$.

Figure~\ref{Fig1} also shows data from computer simulation \cite{6_McSearJac1997} in the absence of a matrix ($\eta_{0} = 0$). Theoretical curves qualitatively reproduce the data from computer simulations but slightly underestimate the value of the fraction $X$, particularly in the nematic region. The results of improving the theory are due to the improvement of the expression (\ref{eq:2.21}) for the contact value of a binary function. Both theoretical and computer data show a jump of $X$ in the area $\eta_{1}$ of the order of $0.345$ (theory) and $\eta_{1} \approx 0.351$ for the simulation. This jump is due to the phase transition from the isotropic phase to the nematic phase. As the matrix packing fraction increases, the value of $X$ decreases and the jump of $X$ shifts toward smaller $\eta_{1}$.

\begin{figure}[h!]
	\begin{center}
		\includegraphics[clip,width=0.6\textwidth,angle=0]{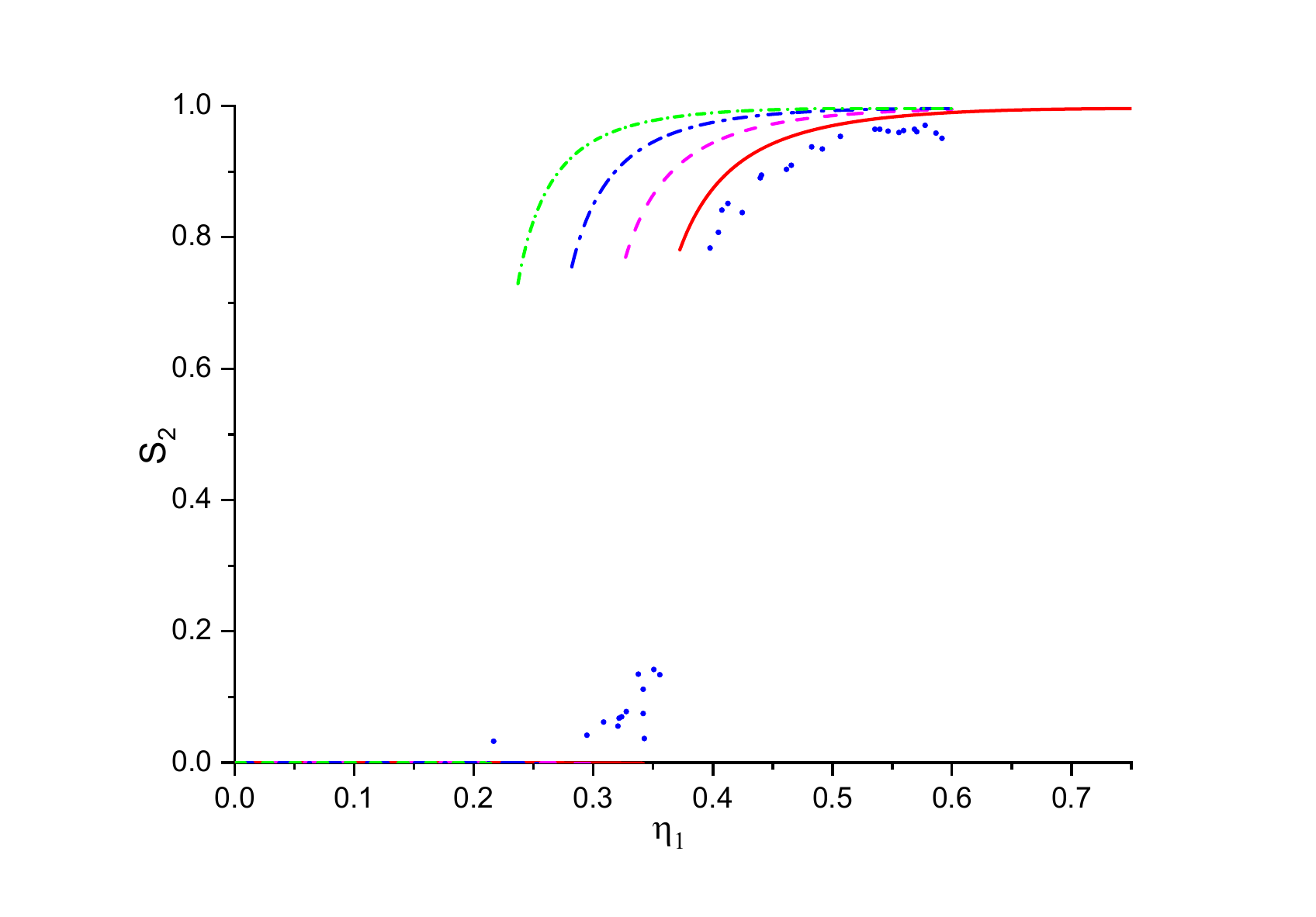}
		\caption{(Colour online) The nematic order parameter $S_{2}$ as a function of $\eta_{1}$ for the dimerization of hard spherocylinders with an aspect ratio of $L_{1}/D_{1} = 5$. Data points represent results from computer simulations  \cite{6_McSearJac1997}, while the curves are based on our theoretical calculations. The colors of the curves indicate various packing fractions of the matrix, represented by $\eta_{0} = 0$ (red color), $0.1$ (magenta color), $0.2$ (blue color), and $0.3$ (green color).
		}
		\label{Fig11}
	\end{center}
\end{figure}

Figure~\ref{Fig11}  shows the dependence of the parameter of nematic order $S_{2}$ for the same system. $S_{2}$ is defined as:
\begin{equation}
	S_{2} = \int f_{1} (\Omega) P_{2} (\cos \theta) \, \rd\Omega,
	\label{eq:3.1}
\end{equation}
where $P_{2}(\cos \theta)$ is the second Legendre polynomial.

\begin{figure}[h]
	\begin{center}
		\includegraphics[clip,width=0.6\textwidth,angle=0]{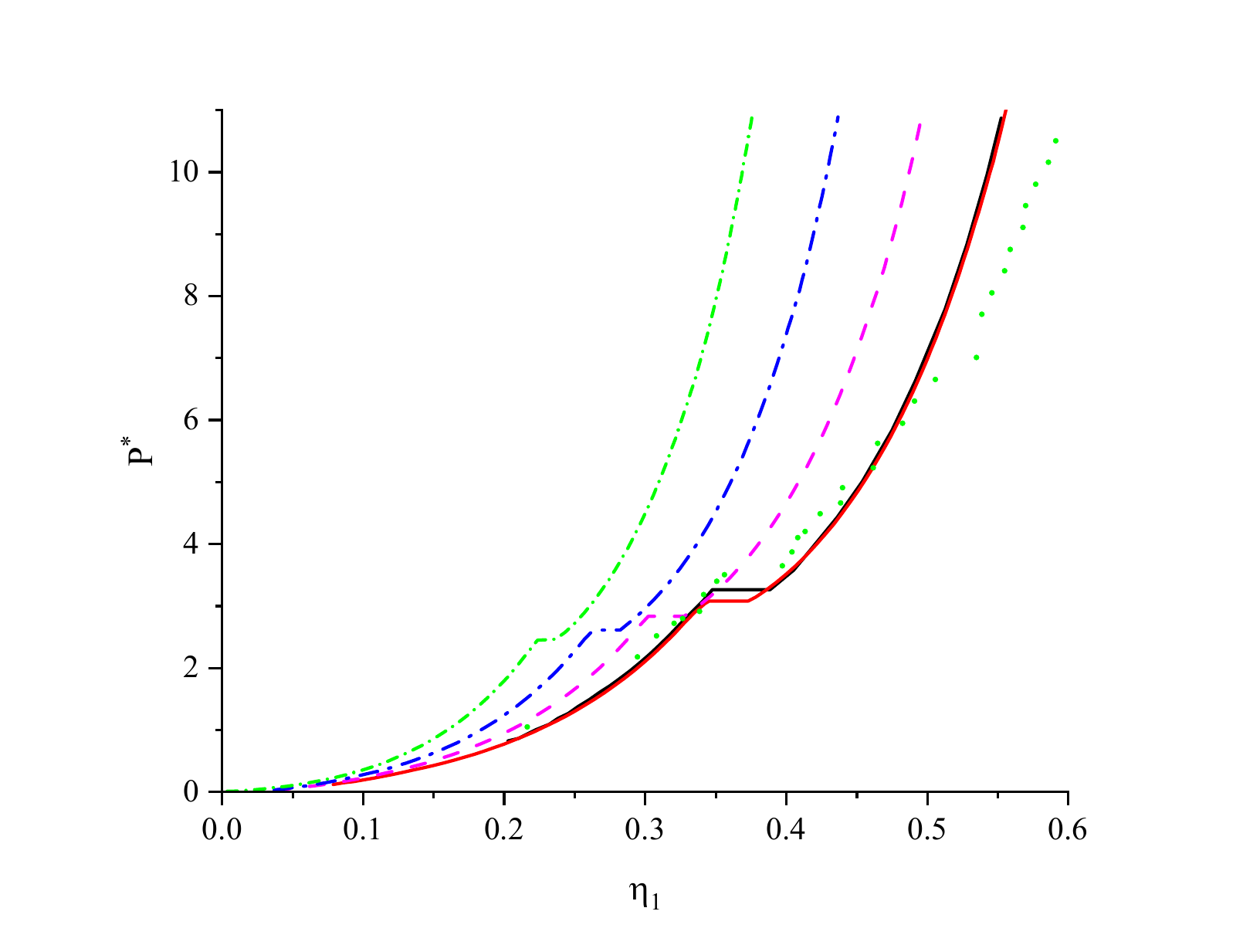}
		\caption{(Colour online) The pressure $P^{*}$ as a function of fluid packing fraction for dimerizing hard spherocylinders with $L_{1}/D_{1} = 5$. The curves represent isotropic and nematic branches obtained from the theory. The colors of the curves indicate various packing fractions of the matrix, represented by $\eta_{0} = 0$ (red color), $0.1$ (magenta color), $0.2$ (blue color), and $0.3$ (green color). The points represent the computer simulation data. The black curve represents the theoretical results for the bulk case from \cite{6_McSearJac1997}.}
		\label{Fig22}
	\end{center}
\end{figure}

As can be seen, the order parameter for the bulk case agrees well with the data from computer simulation \cite{6_McSearJac1997}. In the presence of the matrix, the curves for the order parameter are shifted toward lower densities. The  location of isotropic-nematic transition is  confirmed by the variation of the nematic order parameter $S_{2}$ (figure~\ref{Fig11}) and the fraction of monomer $X$ (figure~\ref{Fig1}) as function of packing fraction $\eta_{1}$.

\begin{figure}[h!]
	\begin{center}
		\includegraphics[clip,width=0.5\textwidth,angle=0]{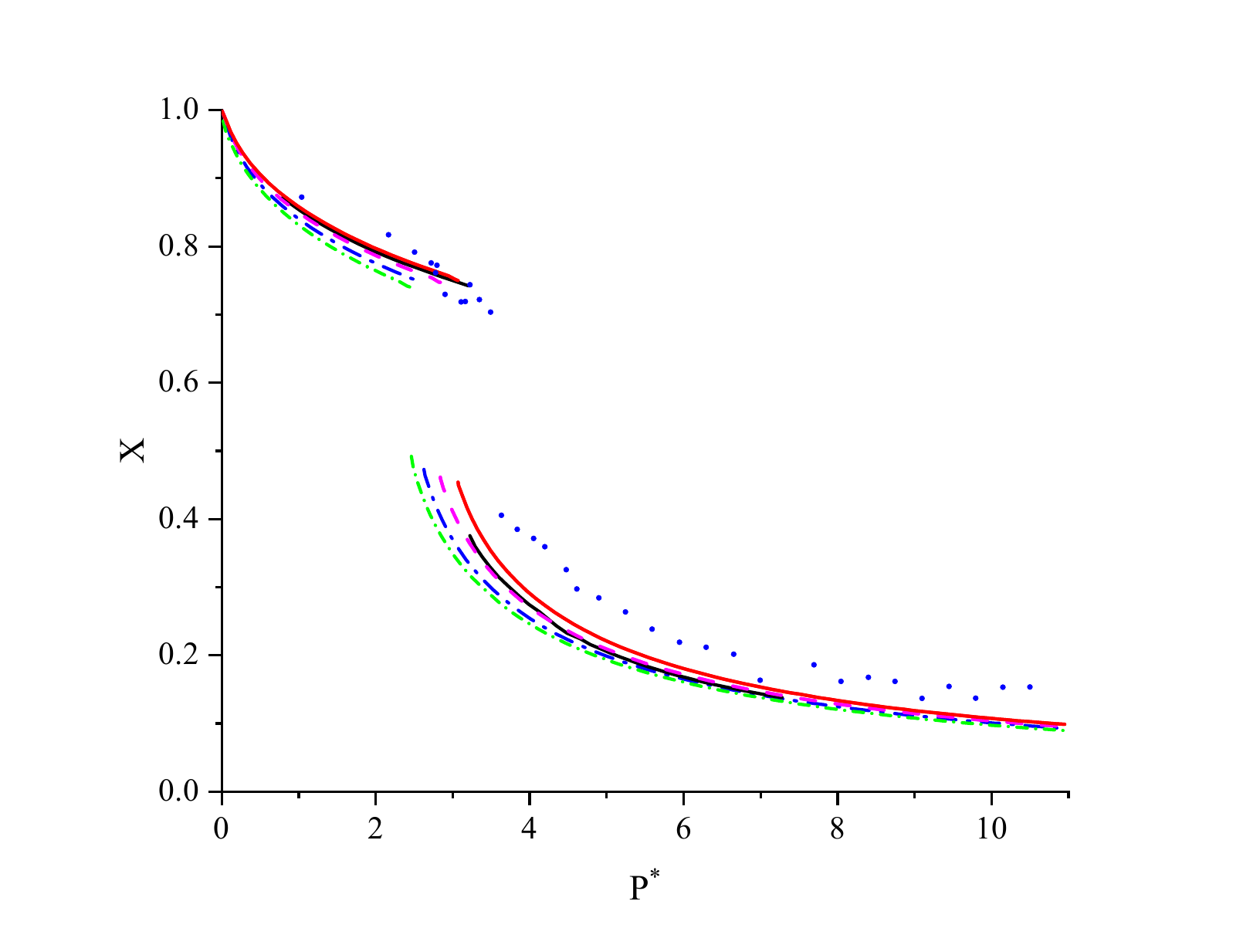}
		\caption{(Colour online) The fraction of monomers $X$ as a function of nondimensional pressure $P^{*}$ for dimerizing hard spherocylinders with $L_{1}/D_{1} = 5$. The curves correspond to isotropic and nematic branches obtained from the theory. The colors of the curves indicate various packing fractions of the matrix, represented by $\eta_{0} = 0$ (red color), $0.1$ (magenta color), $0.2$ (blue color), and $0.3$ (green color). The points represent computer simulation data. The black curve represents the theoretical results for the bulk case from \cite{6_McSearJac1997}.}
		\label{Fig2}
	\end{center}
\end{figure}

Figure~\ref{Fig2}  illustrates the effect of disordered porous media on the fraction of monomers $X$ as a function of pressure $P$ in nondimensional form $P^* = PV_{1}/kT$, where $V_{1} = \frac{1}{6} \piup D_{1}^{3} + \frac{1}{4} \piup L_{1} D_{1}^{2}$ is the volume of a hard spherocylinder. The most striking feature of this figure is the exceptionally clear isotropic-nematic transition for the dimerizing spherocylinders. The degree of association is predicted accurately by the theory in the isotropic and nematic phases, as is its first-order jump at the nematic phase. In the presence of the matrix, the curves are shifted toward lower pressures. Figure~\ref{Fig22} illustrates the effect of porous media on pressure as a function of density for the dimerization of hard spherocylinders. For the bulk case ($\eta_{0} = 0$), the theoretical results qualitatively reproduce the computer simulation data up to the point $\eta_{1}=0.482$, $P^{*}=5.78$ corresponding to the nematic-smectic A transition. This point cannot be  predicted by the considered here Onsager-like theory. The influence of matrix packing fraction $\eta_{0} = 0.1, 0.2, 0.3$ was studied, and the data from computer simulation \cite{6_McSearJac1997} in the absence of a matrix ($\eta_{0} = 0$) are also presented. It is shown that with increasing matrix packing fraction $\eta_{0}$, the isotropic-nematic phase transition shifts to lower densities, with this effect being more noticeable for pressure (figure~\ref{Fig22}). In figure~\ref{Fig2} the theoretical results obtained in \cite{6_McSearJac1997} are also shown by black curves. As we can see, there is a good agreement between our theoretical results and theoretical results obtained in \cite{6_McSearJac1997} only in the isotropic phase. However, in the nematic phase, our theoretical results are in better agreement with simulation data. This is not surprising because in \cite{6_McSearJac1997}, a trial function was used for the description of the singlet distribution functions $f_{1}(\Omega_{1})$ and $f_{10}(\Omega_{1})$. As known \cite{VrLek}, the trial function approach overestimates the nematic ordering at the isotropic-nematic transition.

\begin{figure}[h!]
	\begin{center}
		\includegraphics[clip,width=0.65\textwidth,angle=0]{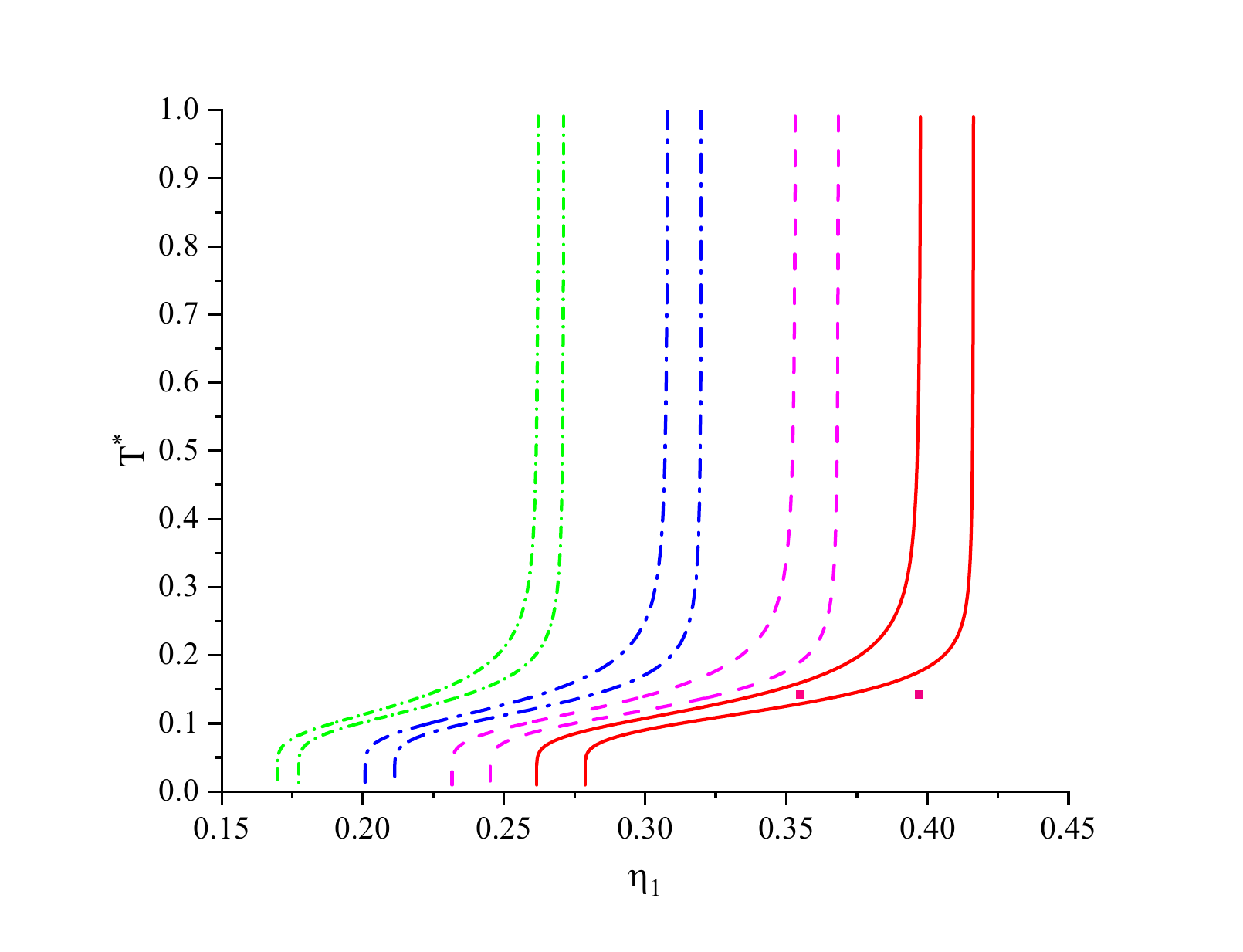}
		\caption{(Colour online) The temperature dependence of the isotropic-nematic phase transition obtained from the theory for dimerizing hard spherocylinders with $L_{1}/D_{1} = 5$. The reduced temperature is defined as $T^{*} = kT/\epsilon_{1}$. The red lines correspond to the bulk case, while other colors indicate cases with different packing fractions of the matrix $\eta_{0}$ (similar as in the previous figures). The left-hand branch corresponds to the end of the stable isotropic phase, and the right-hand branch to the beginning of the stable nematic phase. The points are taken from computer simulation \cite{6_McSearJac1997}.}
		\label{Fig3}
	\end{center}
\end{figure}

Finally, figure~\ref{Fig3} presents the phase diagram of the isotropic-nematic phase transition. The phase transition predicted by the theory, as illustrated in figure~\ref{Fig3} by lines, is qualitatively consistent with the results of Monte Carlo modelling \cite{6_McSearJac1997}. At high temperatures, the transition densities correspond to those expected in a system of simple monomers without dimerization. At very low temperatures, where dimers predominate, the transition density is much lower and exhibits little sensitivity to temperature changes. Dimerization stabilizes the nematic phase relative to the isotropic fluid. Only within a narrow temperature range does the density change rapidly. In this region, dimerization greately affects the density ranges near the isotropic-nematic (I--N) transition zone, where there is also a noticeable expansion of the two-phase area. The theoretical conclusions summarized in figure~\ref{Fig3} indicate that the reduced temperature $T^{*} = kT/\epsilon_{1} = 0.1429$ corresponds to the region where dimerization will have the greatest influence on the I--N phase transition. The presence of a porous medium leads to a shift in the phase transition density to lower values.

\section{Conclusions}

In this article, we have presented a study on phase transitions in the system of dimerizing hard spherocylinders taking into account the influence of a porous medium. The porous medium is considered as quenched configurations of randomly distributed hard spheres. The developed theoretical approach is based on the combination of the recently developed in our previous works scaled particle theory for the description of a hard spherocylinder fluid in random porous media \cite{HolShmot2018,book_11}, with Wertheim multidensity formalism \cite{CM_2015_54,CM_2015_54_a} for the treatment of the effects of dimerization of spherocylinders. We note that in order to check the accuracy of our theoretical predictions, we compare the obtained theoretical results with the data from computer modelling \cite{6_McSearJac1997}  for the dimerization model of hard spherocylinders in the bulk case. However we should note that computer simulations for dimerazing model of hard spherocylinder fluid in porous media are needed for subsequent progress. Using a combination of theoretical analysis and computer modelling, we were able to gain a deeper understanding of how associative interactions and properties of the porous matrix affect the isotropic-nematic phase transition. It was discovered that dimerization effectively promotes the stabilization of the nematic phase. This is confirmed by comparing theoretical predictions with computer modelling data for a bulk case. The study showed that the porous medium causes a shift in the phase transition to lower density values. Increasing the packing fraction of the matrix enhances this effect. The analysis of temperature dependencies indicates a significant influence of dimerization on the phase transition, especially at low temperatures where dimers predominate. This research not only expands our understanding of the fundamental aspects of phase transitions in colloidal systems but can also  provide new prospects for the development of materials with specific properties.  Future studies will focus on exploring the impact of more complex particle shapes and a variety of associative interactions on the system behavior. We should emphasize that Wertheim extended his formalism to account for polymerization for description of a flexible chain of hard spheres~\cite{CM_2015_55,CM_2015_55_a,Wertheim1987}. We note that in  \cite{DeMichele2012,DeMichele2019} the phase behaviour of hard spherocylinder fluid with possibility of polymerization between spherocylinders were considered but without the presence of porous media. Thus, the possibility for investigation of polimerizable hard spherocylinder fluid in presence of porous media is possible. However, we should note that in the framework of the considered approach, the orientational ordering is connected only with a nonspherical form of colloids. Associative interactions stabilize the ordering but cannot be the source of ordering due to flexibility of clustering in Wertheim formalism. Another important aspect of generalization
of the considered theory can be connected with introduction of stiffness parameters in the framework of the approach developed by Binder group \cite{Egorov2016}.

\section*{Acknowledgement}
We gratefully acknowledge financial support from the
National Research Foundation of Ukraine (project No. 2020.02/0317).
The authors express their gratitude to Taras Patsahan  for useful discussions.

\ukrainianpart

\ukrainianpart

\title{Димеризаційні тверді сфероциліндри в пористому середовищі}
\author{В. І. Шмотолоха, М. Ф. Головко}
\address{
 Інститут фізики конденсованих систем Національної академії наук України,  вул.~Свєнціцького 1, 79011 Львів, Україна,
}

\makeukrtitle

\begin{abstract}
	\tolerance=3000%
	Це дослідження зосереджується на унікальній фазовій поведінці несферичних плямистих колоїдів у порис\-тих середовищах. На основі теорії масштабної частинки (ТМЧ) було вдосконалено методику та застосовано її для аналізу термодинамічних властивостей несферичних плямистих частинок у невпорядкованому пористому середовищі. Використовуючи асоціативну теорію рідин в поєднанні з ТМЧ, ми дослідили вплив асоціативних взаємодій та з'єднання між функціональними вузлами частинок на процес формування нематичної фази. Було проведено розрахунки орієнтаційних та просторових розподілів, які допомогли зрозуміти фазову поведінку частинок при переході від ізотропної до нематичної фази в умовах просторового обмеження, створеного невпорядкованою матрицею пористого середовища.
	
	\keywords плямисті колоїди, сфероциліндри, димеризація, невпорядковані пористі середовища, геометрична пористість, пористість пробної частинки
	
\end{abstract}

\end{document}